# Transcranial Focused Ultrasound for BOLD fMRI Signal Modulation in Humans

Leo Ai, Jerel K. Mueller, Andrea Grant, Yigitcan Eryaman, and Wynn Legon

*Abstract*— **Transcranial focused ultrasound (tFUS) is an emerging form of non-surgical human neuromodulation that confers advantages over existing electro and electromagnetic technologies by providing a superior spatial resolution on the millimeter scale as well as the capability to target sub-cortical structures non-invasively. An examination of the pairing of tFUS and blood oxygen level dependent (BOLD) functional MRI (fMRI) in humans is presented here.**

## INTRODUCTION

Transcranial focused ultrasound (tFUS) is a new and promising non-surgical low-energy technique for inducing transient modulation with high spatial resolution, adjustable focus and low tissue attenuation. Ultrasound can noninvasively stimulate the hippocampus and motor cortex of intact mice [1, 2], modulate monosynaptic and polysynaptic spinal reflexes in cats [3] and disrupt seizure activity in cats [4], rats [5], and mice [6]. In addition, tFUS has been used safely and effectively for intact neural stimulation in mouse [7], rabbit [8], sheep [9] and monkey [10] and our work has shown tFUS to also be a safe and effective method of transient transcranial cortical stimulation in humans [11, 12]. Using EEG, our work has shown tFUS to inhibit the amplitude of somatosensory evoked potentials (SEP) and to affect the power and phase of beta and gamma frequencies [13] that has since been expanded upon by other groups [14], demonstrating ultrasound as an efficacious form of highly focal transient stimulation for use in humans. Because of tFUS' high spatial resolution and ability to stimulate at depth, it is of particular interest to pair tFUS with magnetic resonance imaging (MRI) to take advantage of its superior whole brain resolution. Previous literature has shown in craniotomized rabbits that 3.3 W/cm$^2$ spatial-peak pulse-average intensity ($I_{sppa}$), 0.69 MHz focused ultrasound to the sensorimotor cortex resulted in a detectable BOLD response that corresponded well in space to the FWHM sonication pressure [8]. Here we report on the feasibility of performing concurrent tFUS/MRI in humans, the issues inherent in this process and the results from our experiments examining cortical BOLD response at 3T and sub-cortical BOLD response at 7T.

## CHARACTERIZATION AND APPLICATION OF tFUS WAVEFORMS

We used two separate focused transducers for this study. For the 3T experiment, we used a 0.5MHz transducer that had an active diameter of 60 mm and focal length of 55 mm producing a focal FWHM intensity volume of 48.64 mm$^3$. For the 7T experiment we used a 0.86 MHz focused transducer that had an active diameter of 64 mm and a focal length of 54 mm producing a focal FWHM volume of 35.77 mm$^3$. Transcranial ultrasonic neuromodulation waveforms were generated using a two-channel, 2-MHz function generator (BK Precision Instruments, CA, USA). Channel 1 was set to deliver US at a pulse repetition frequency of 1.0 KHz, and channel 2 was set to drive the transducer at the acoustic frequency in a bursting mode, with channel 1 serving as an external trigger for channel 2. The pulse duration of the waveform was set by adjusting the number of cycles per pulse on channel 2. Stimulus duration was set by adjusting the number of pulses on channel 1. The output of channel 2 was sent through a 100W linear RF amplifier (E&I 2100L; Electronics & Innovation) matched to the impedance of the transducer by an external matching network. The waveforms were as described in Legon et al. 2014 [12]. For the 3T experiment,

L. Ai is with the Department of Physical Medicine and Rehabilitation, University of Minnesota, Minneapolis, MN 55455 USA (e-mail: leoai@umn.edu)

J. K. Mueller is with the Department of Physical Medicine and Rehabilitation, University of Minnesota, Minneapolis, MN 55455 USA (e-mail: jkmuelle@umn.edu)

A. Grant is with the Center for Magnetic Resonance Research, University of Minnesota, Minneapolis, MN 55455 USA (e-mail: gran0260@umn.edu)

Y. Eryaman is with the Center for Magnetic Resonance Research, University of Minnesota, Minneapolis, MN 55455 USA (e-mail: yigitcan@umn.edu)

W. Legon is with the Department of Physical Medicine and Rehabilitation, University of Minnesota, Minneapolis, MN 55455 USA (phone: 612-626-1183; e-mail: wlegon@umn.edu).

180 cycles of the 0.5 MHz acoustic frequency were pulsed at 1 KHz resulting in a 36% duty cycle. 500 cycles of the pulse repetition frequency were delivered resulting in a 500 ms duration stimulus. For the 7T experiment, 420 cycles of the 0.86 MHz acoustic frequency (50% duty cycle) were pulsed at 1 KHz with the same 500 ms duration as above. Stimulus timing was controlled by custom written Matlab (Mathworks, MA, USA) scripts timed according to TR onset of the MRI scanner. For transducer mounting on the head, the hair was parted to expose the scalp, and the transducer thoroughly coated with acoustic coupling gel and secured in place using surgical meshing and athletic pre-wrap. For the 3T experiment, the transducer was offset from the head 3cm and coupled using a coupling cone filled with acoustic gel. Positioning of the transducer was performed using a stereotaxic neuronavigation system (Rogue Research Inc., CAN) that has been adapted to work with the ultrasound transducers.

Ultrasound acoustic intensity profiles from the above waveforms were measured in degassed, deionized free water and through hydrated human cranium in a 300L acoustic test tank (Precision Acoustics, UK) using a calibrated hydrophone (HNR-0500, Onda Corporation, CA). Scans were first performed without bone and then performed using a 6-mm-thick fragment of human parietal bone (degassed and rehydrated for 48 h) between the transducer and the hydrophone. From this data, we adjusted the driving voltage for each transducer to produce a spatial peak pulse average intensity $I_{sppa}$ after bone transmission of $6W/cm^2$.

# tFUS with MRI

## tFUS Inducing BOLD Signal at 3T

Six neurologically healthy volunteers (20 – 36 years) were scanned at the Virginia Tech Carilion Research Institute on a Siemens (Erlangan, Germany) Tim Trio 3T scanner using a 12 channel head matrix coil. Prior to scanning, volunteers were fitted with the transducer targeted at the primary motor cortex hand knob of the dominant hemisphere. Prior to the functional scans, a T1-weighted anatomical image was acquired to align with each subject's BOLD contrast data. Functional images of BOLD contrast signals were acquired using a gradient echo echo-planar imaging (GRE EPI) sequence (TR = 2000 ms, TE= 30 ms, flip angle = 90, FOV = 190 mm, slices = 33, slice thickness = 3 mm, 450 images). An event-related design was employed where 90 0.5 sec pulsed ultrasound stimuli were delivered every 6-7 TRs (12-14 secs). The resulting time courses were analyzed using Statistical Parametric Mapping (SPM8; www.fil.ion.ucl.ac.uk/spm/). The functional data was slice-time corrected, realigned, preprocessed by linear trend removal, temporal high-pass filtered (128 sec) and three-dimensional motion corrected using a trilinear interpolation. Functional data sets were transformed into MNI space and co-registered with anatomical data for each subject. The resulting time courses were filtered using an 8-mm Gaussian kernel at full width half-maximum (FWHM). Statistical analysis was performed by fitting the signal time course of each voxel using the general linear model (GLM). The onset of stimulation was used as a regressor and modeled using the canonical hemodynamic response function. Additional regressors included head motion parameters generated during fMRI preprocessing. Individual analysis was for US on/off using one-sample t-tests. Testing for significance at the group level (N = 6) was conducted for condition US on/off using a one-sample t-test. Both whole-head and region of interest (ROI) analysis were conducted. The ROI included the precentral gyrus to the depth of the central sulcus as well as the post-central gyrus to the depth of the central sulcus as determined from the T1 scans. Data is presented at p = 0.001 uncorrected with a cluster threshold of 5 voxels. There was no statistically significant area of activation on the whole head group level or on the group level ROI of the sensorimotor region. Further exploration of individual subject data revealed that 3 of the 6 participants showed no discernable BOLD response. However, the remaining 3 participants' revealed very focal BOLD response in the sensorimotor region in good accordance with the focus of the ultrasound beam (Figure 1). The lack of effect for 3 of the 6 participants is the likely reason for a lack of group effect.

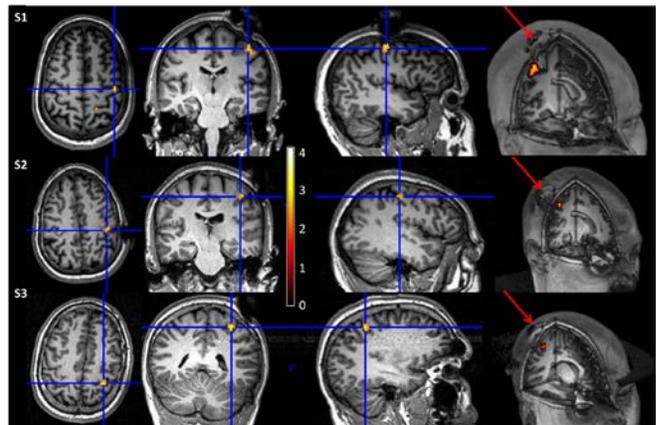

Figure 1. Results from 3/6 subjects showing clear BOLD activity as a result of tFUS. Arrows on right side showing contrast of transducer on the head. Scale is T-values.

*tFUS Inducing BOLD Signal at 7T*

A pilot study was performed to examine the feasibility of tFUS to induce a BOLD fMRI signal in subcortical tissue. The study was performed on a neurologically intact volunteer and imaged on a 7T Siemens magnet using a Nova Medical 1x24 Head Coil (Wilmington, MA, USA) at the University of Minnesota's (UMN) Center for Magnetic Resonance Research (CMRR). The location of the transducer placement was determined using the BrainSight neuronavigation system and targeted the left head of caudate.

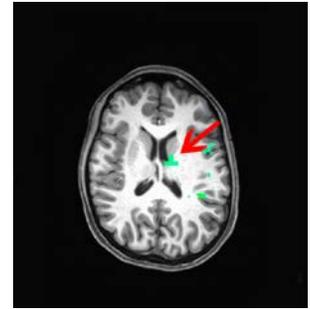

Figure 2. Generated BOLD signal in the caudate at 7T

The tFUS stimulation was delivered with a block design (5 off/on cycles, stimulation delivered every 6 TRs during on cycles). A GE EPI sequence was used to acquire the functional data (parameters: TR = 600ms, TE = 21.4ms, flip angle = 40, voxel dimensions = 2x2x2mm, 1000 images), along with a T1 anatomical scan. The functional data was analyzed with Analysis of Functional NeuroImages (AFNI) [15] Correlation analysis was used to analyze the fMRI data using a reference function generated by convolving the stimulation paradigm with the hemodynamic response function. As part of the analysis process, motion correction was performed to minimize motion effects, and a Gaussian filter (FWHM = 4mm) was applied for smoothing. A Z score threshold of 3.5 ($p = 0.0002$, uncorrected) was applied to isolate activations. Initial results indicate a focal BOLD response in the targeted caudate area (Figure 2).

## MR COMPATIBILITY – TEMPERATURE

One of the concerns with concurrent tFUS/MRI is transducer and cable heating. A study was performed to examine surface temperature changes of the transducer operated on a 7T Siemens magnet using a Nova Medical 8x32 Head Coil at UMN's CMRR. A spin echo pulse sequence was used to image the transducer over a period of 20 minutes, with temperatures being recorded, to simulate data acquisition using parameters set to reach the SAR limits set by the FDA (3.2W/kg per 10-minute exposure) [16]. The temperature tests were run within the scanner with the transducers unpowered and powered to deliver a mechanical index of 1 using 0.5 sec pulsed ultrasound stimulation delivered every 5 seconds for the duration of the 20 minute temperature recording session. Two 0.5MHz transducers were used for this examination, one with a diameter of 64 mm and the second with a diameter of 30mm. The transducer faces were first filled with ultrasound gel, and fiber optic temperature probes (Lumasense Technologies, CA, USA) were then placed on the center and along the edges of the transducer faces inside of the ultrasound gel. The transducers, with the temperature probes attached, were then secured to an agar phantom to simulate a real data acquisition. The results with the transducers running showed that the center face recorded the greatest increase in temperature of approximately 1° C for the smaller transducer and 0.75° C for the larger transducer over the recorded period of 20 minutes. In the off state, no temperature increases were detected. One degree centigrade is an acceptable surface temperature increase, indicating minimal risk of surface heating using tFUS/MRI at high field (7T).

## MR COMPATIBILITY - ARTIFACTS

Typical ultrasound transducers contain lead zirconate titanate (PZT) for the piezoelectric element. This is a paramagnetic inorganic compound that produces susceptibility artifacts in the MRI scanner, a potentially significant problem that needs to be resolved and needs to be taken into consideration during transducer manufacturing, especially at high field strengths. Using custom designed PZT ceramic focused transducers with normal and half the normal amount of PZT, MR artifact tests were performed on an agar phantom on a 7T Siemens (Erlangan, Germany) magnet using a Nova Medical 8x32 Head Coil (Wilmington, MA) and a GRE EPI pulse sequence (at UMN's CMRR). The normal PZT element produced considerable magnetic susceptibility artifact. The PZT element was then translated in 1cm steps away from the phantom to assess the effect on the artifact. At 4cm from the agar phantom, the artifact resolved (Figure 3).

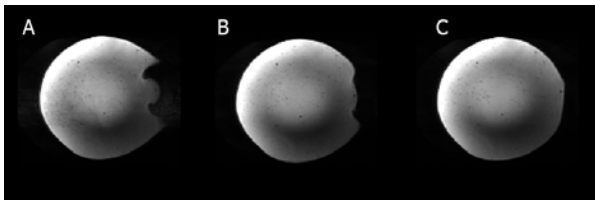

Figure 3. MR susceptibility artifact seen with PZT element translated away from an agar phantom. A: PZT up against the phantom. B: PZT 1 cm from phantom. C: PZT 4cm from

It should be noted that even with a transducer containing half the PZT, the susceptibility artifacts are not eliminated, but can be reduced to a state that could be acceptable for studies interested in examining cortical activity directly under the transducer with careful shimming (Figure 4). However the issue of susceptibility artifacts still can prohibit studies that involve examinations closer to the surface of the cortex, and more work will have to be done to further reduce artifacts for such studies.

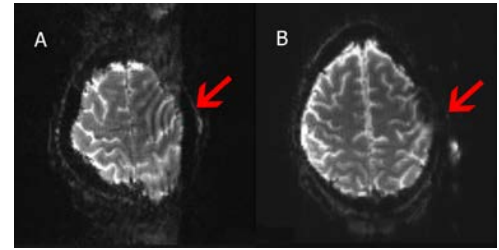

Figure 4. A: MR susceptibility artifact seen before being shimming, and B: after shimming. Note transducer location in B from vitamin E pill. Arrows point to artifact caused by transducer.

## FURTHER DISCUSSION

The results from the preliminary studies indicate that it is possible to detect BOLD fMRI signals from tFUS at both 3T and 7T field strengths in both cortical and subcortical regions of the brain. BOLD activations were detected in the M1/S1 area in the 3T studies, and a BOLD activation was detected in the caudate in the 7T study. There are considerable potential issues with MR artifacts and heating from the transducer, though it looks like these issues can be overcome with proper transducer design.

The 3T study presented results with BOLD fMRI activity detected in the targeted M1/S1 region. The results were not consistent where 3/6 participants showed no discernable BOLD activation. The reasons for this are not currently clear and need to be elucidated. This could be related to the amount of energy, the pulsing strategy, the event design and perhaps an under-powered ability to detect the signal. This study used an event-related design with a long inter-stimulus interval that may not be efficacious for detecting response. We used an ultrasound waveform that delivered equivalent intensities to the brain as reported by [8] though with differences in pulsing strategy and duty cycle. It is possible that these differences lead to our variability and should be examined in future studies. There is also the consideration that ultrasound does not directly affect neuronal kinetics but rather affects local microvasculature through mechanical perturbation that leads to the BOLD response. We cannot be definitive that this is not the case but, the fact that ultrasound results in behavioral improvements [12, 14, 17] suggests that ultrasound is mediating a neuronal response.

The 7T pilot study also did present with a BOLD fMRI activation in the targeted caudate area, but there are steps that can be taken to further strengthen the case for deep brain stimulation in humans with tFUS. The acoustic frequency used was not optimal for bone transmission [18], but was used to achieve a better focal volume for specific targeting of small deep-brain structures. Transmission loss can be overcome however by delivering a high initial intensity outside of the head and we have shown that the skull does not serve to distort the acoustic field and may in some cases serve to help focus it [12].

These initial studies provide evidence that detection of a BOLD response is achievable non-invasively with tFUS in humans. However, we realize that optimization is necessary. MRI parameters need to be further examined to maximize the ability to reliably detect tFUS induced BOLD signals, especially shimming strategies and pulse sequence selection (i.e., GRE vs spin-echo sequences). Magnetic susceptibility artifacts caused by placing the transducer in the MR environment will continue to be a concern for all continuing studies involving pairing tFUS with MRI based techniques. It was found that careful shimming can reduce these artifacts to acceptable levels, but this may not always be the case. Transducer materials and design have the largest effect on susceptibility artifacts and there are examinations currently underway to determine the desired transducer characteristics that would be the most beneficial for operating in the MR environment. Once the optimal transducer characteristics are determined, the optimal MRI parameters and procedures will also be thoroughly examined to maximize signal detectability.

## CONCLUSION

The studies performed would indicate that it is possible to induce BOLD fMRI signals using tFUS targeting cortical and subcortical structures at 3T and 7T respectively. Optimizations in ultrasound pulse parameters, MRI data acquisition parameters, and ultrasound transducer design needs to be made, but these preliminary studies show promise in using tFUS with BOLD fMRI, offering potentially many advantages over currently used neuromodulation techniques. The tFUS technique is able to target sub-cortical structures with a precision on the millimeter scale. As such, it is of utmost importance to the development of tFUS to examine its effects on BOLD fMRI. Observations from such examinations could have important implications in global brain mapping efforts and in many forms of guided therapies in clinical settings.

## ACKNOWLEDGMENTS

The authors would like to thank Dr. Jamie Tyler and the Virginia Tech Carilion Research Institute for use of their scanners, and Dr. Essa Yacoub at CMRR for his expertise.